\documentclass[twocolumn,showpacs,preprintnumbers,prl]{revtex4}
\usepackage{graphicx}
\usepackage{dcolumn}
\usepackage{bm}
\usepackage{color}

\begin{document}

\title{Purification of Single Photons from Room-Temperature Quantum Dots}
%\title{Room-temperature solid-state single-photon source with high purity and controllable waveforms}
%\title{Room-temperature quantum-dot-based single-photon source with high purity and controllable waveforms}
%\title{High-purity waveform-controlled single photons from a single room-temperature quantum dot}
%\title{Purification of temporally long single photons from a room-temperature quantum dot}
\author{Shih-Wen Feng}
\author{Chun-Yuan Cheng}
\author{Chen-Yeh Wei}
\author{Jen-Hung Yang}
\author{Yen-Ru Chen}
\author{Ya-Wen Chuang\footnote[2]{Present address: Department of Physics, The Pennsylvania State University, USA.}}
\author{Yang-Hsiung Fan\footnote[3]{Present address: Ezconn Corporation.}}
%\author{Chao-Yuan Wang}
\author{Chih-Sung Chuu\footnote[1]{cschuu@phys.nthu.edu.tw}}
%\email{cschuu@phys.nthu.edu.tw}
\affiliation{Department of Physics, National Tsing Hua University, Hsinchu 30013, Taiwan\\
and Frontier Research Center on Fundamental and Applied Sciences of Matters, National Tsing Hua University, Hsinchu 30013, Taiwan}

\begin{abstract}

Single photon emitters are indispensable to photonic quantum technologies. Here we demonstrate waveform-controlled high-purity single photons from room-temperature colloidal quantum dots. The purity of the single photons does not vary with the excitation power, thereby allowing the generation rate to be increased without compromising the single-photon quality.

\end{abstract}

\pacs{03.67.Bg, 42.65.Lm, 42.50.Dv}

\maketitle

Single-photon emitters (SPEs) are vital to photonic quantum technologies \cite{Kimble08, O'Brien09, Northup14}, with examples ranging from quantum communication and computation to quantum metrology. Among a variety of SPEs available today, semiconductor quantum dots are promising due to their outstanding optical properties and scalability \cite{Shields07}. While much effort has been devoted to the self-assembled quantum dots at cryogenic temperatures, single-photon emission has also been demonstrated with the room-temperature colloidal core/shell quantum dots~\cite{Michler00b}. The colloidal quantum dots exhibit high quantum yield and photostability at room temperature. However, due to the spectrally broad and inseparable biexciton spectrum at room temperature, the single-photon purity of these SPEs is incomparable to their cryogenic-temperature counterparts and may limit their applicability. 

In this Letter, we demonstrate the purification of single photons to realize high-purity single photons from the room-temperature colloidal quantum dots. The purification is made possible by the long temporal length of the single photons \cite{Michler00b, Lounis00} and the waveform engineering that eliminates the biexciton emission. The waveform engineering not only enables the purification but also preserves the single-photon purity at high excitation power, thereby allowing the generation rate to be increased without compromising the single-photon quality. By properly shaping the single photons, we achieve a single-photon purity of $g^{(2)}(0)=0.01$ which does not vary with the excitation power or between different quantum-dot samples.

There is a distinction between our waveform engineering and that achieved with the single trapped ions \cite{Keller04} or cold atomic ensembles \cite{Balic05, Du08, Kolchin08, Chen10, Zhao15}. While our single photons have a temporal length comparable to those from the single trapped ions and cold atomic ensembles, their bandwidths are dominated by dephasing and not lifetime limited. Thus, the waveform engineering demonstrated here only controls the temporal envelope of the single photons. On the contrary, the temporal wave functions of the single photons can be coherently controlled in the single trapped ions or cold atomic ensembles.

\begin{figure}[t]
\centering
\includegraphics[width=1 \linewidth]{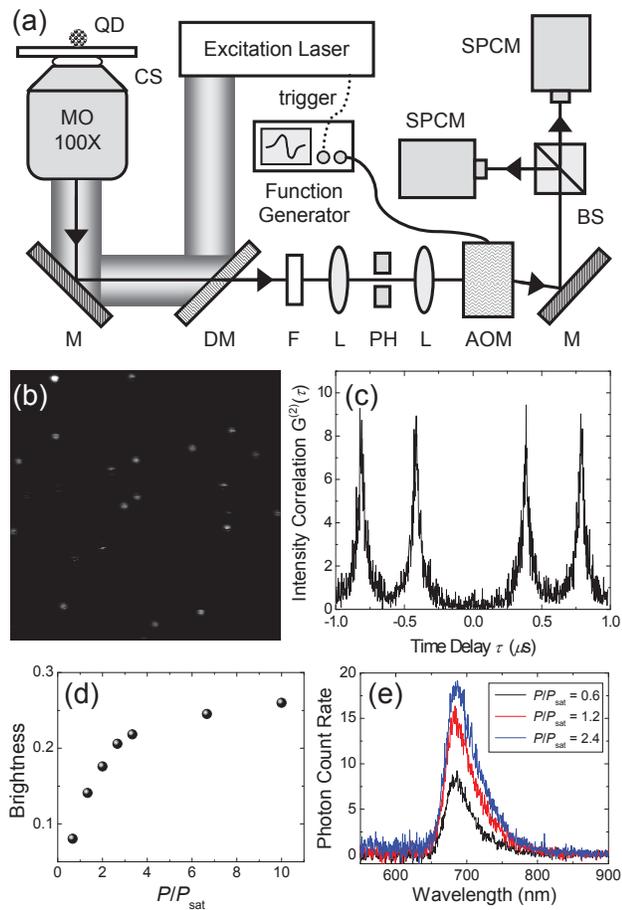}
\caption{\label{fig:1} (color online) Single photons from room-temperature quantum dots. (a) Single quantum dots (QDs) are spin coated on a cover glass (CS) and optically excited by focused laser pulses using a 100X microscopic objective (MO). The emitted single photons are collected by the same MO, separated from the laser pulses at the dichroic mirror (DM), and detected by the single photon counting modules (SPCMs). To generate waveform-controlled single photons, an acousto-optic modulator (AOM) driven by a function generator is used to modulate the waveform of the single photons. Other optical elements shown are mirror (M), filter (F), pinhole (PH), lens (L) and beam splitter (BS). (b) Photoluminescence image ($20 \times 20 \mu {\rm m}$) of single quantum dots. (c) Antibunching in the Hanbury-Brwon-Twiss experiment. (d) Brightness as a function of excitation power $P$ normalized to the saturation power $P_{\rm sat}$. (e) Spectra of a single quantum dot.}
\end{figure}

The schematic of our experimental setup is illustrated in Fig.~\ref{fig:1}(a). Room-temperature CdSeTe/ZnS quantum dots (7.5-nm radius, 705-nm peak emission) are prepared on a cover slip with a density $< 1~\mu$m$^{-2}$ by spin coating. A single quantum dot is identified by the photoluminescence imaging [Fig.~\ref{fig:1}(b)] and the antibunching feature of its emitted photons in the Hanbury-Brown-Twiss measurement [Fig.~\ref{fig:1}(c)]. The excitation laser, a 405-nm pulsed laser (63-ps pulse width, 1 MHz repetition rate), is tightly focused onto the single quantum dot with a beam waist of 500 nm using an oil-immersion microscope objective (1.4 NA, 100X). The single photons from the quantum dot are collected through the same objective and separated from the excitation beam using a dichroic mirror and two long-pass filters. In addition, confocal microscopy is exploited to increase the signal-to-noise ratio. The detection efficiency of the single photons, including the overall collection efficiency of the optical system (20\%) and the quantum efficiency of the single photons counting module (60\%), is 12\%. The measured brightness (namely, the photon number collected by the first optics per excitation pulse) and spectra of a single quantum dot are shown in Figs.~\ref{fig:1}(d) and \ref{fig:1}(e), respectively, for various normalized excitation power.

%\textbf{Waveform manipulation of the single photons.} The schematic of our experimental setup is illustrated in Fig.~\ref{fig:1}(a). Room-temperature quantum dots are prepared on a cover slip with a density of $< 1~\mu$m$^{-2}$ (see Methods). Single quantum dots are identified by the photoluminescence imaging [Fig.~\ref{fig:1}(b)] and verified, after optically excited, by the antibunching feature of the emitted single photons in the Hanbury-Brown and Twiss measurement [Fig.~\ref{fig:1}(c)] (see Methods). The excitation laser is a 405-nm pulsed laser (1-MHz repetition rate, 1-ps pulse width), which is tightly focused onto the single quantum dot with a beam waist of 700 nm using a oil-immersion microscope objective (1.4 NA, 100X). The single photons from the single quantum dot are collected through the same objective and separated from the excitation beam using a dichroic mirror and two long-pass filters. In addition, confocal microscopy is exploited to increase the signal-to-noise ratio. The total detection efficiency of the single photons, including the collection efficiency of the optical system and the quantum efficiency of the single photons counting modules (Excelitas, SPCM-AQRH-14-FC), is 12 \%.

\begin{figure}[t]
\centering
\includegraphics[width=0.9 \linewidth]{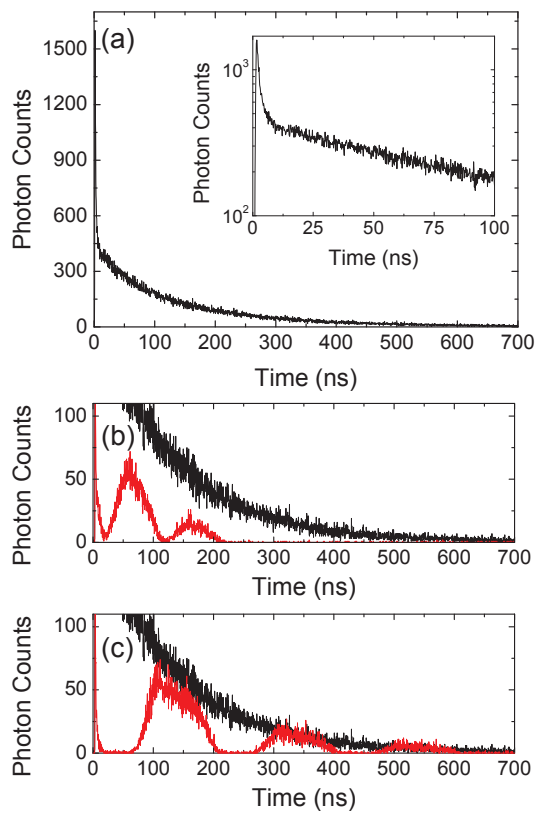}
\caption{\label{fig:2} (color online) Acousto-optic modulation of single photons. (a) Waveform of the unmodulated single photons from a single room-temperature quantum dot. The inset shows the biexponential shape. (b) Waveforms of the single photons modulated by a biased sinusoidal function of frequency 10~MHz (red) and without modulation (black). (c) Waveforms of single photons modulated by a biased square function with a period of 200~ns (red) and without modulation (black). The excitation power is $5.8 \ P_{\rm sat}$.}
\end{figure}

Figure~\ref{fig:2}(a) shows the waveform of the single photons obtained by the time-resolved photoluminescence of a single quantum dot at the single photon level. The waveform is biexponential as evident in the logarithmic scale (inset). The fast decay, with a time constant of $\tau_1 \sim 2$~ns, is contributed mainly by the biexciton emission \cite{Fisher05}. The slow decay, on the other hand, results from the radiative single-exciton transition \cite{Brokmann04} and determines the temporal width of the single photons, $\tau_2 \sim 138$ ns. Such a long temporal length allows us to manipulate the waveform of the single photons with an acousto-optic modulator. 

To control the waveform, special care has to be taken to ensure that the front edge of the single photon arrives at the acousto-optic modulator at the same time as that of the electric signal driving the acousto-optic modulator. For this purpose, the excitation laser is triggered by the function generator (after a controlled time delay) which also drives the acousto-optic modulator. Figure~\ref{fig:2}(b) shows the waveform of the single photons modulated by a biased sinusoidal function of frequency 10 MHz. The modulation results in two visible periods of a sinusoidal wave with each width of 100 ns. Figure~\ref{fig:2}(c) shows another example of shaped single photons which is modulated by a biased square function with a period of 200 ns. These examples demonstrate the feasibility of shaping the single photons from the room-temperature quantum dots.

\begin{figure}[t]
\centering
\includegraphics[width=1 \linewidth]{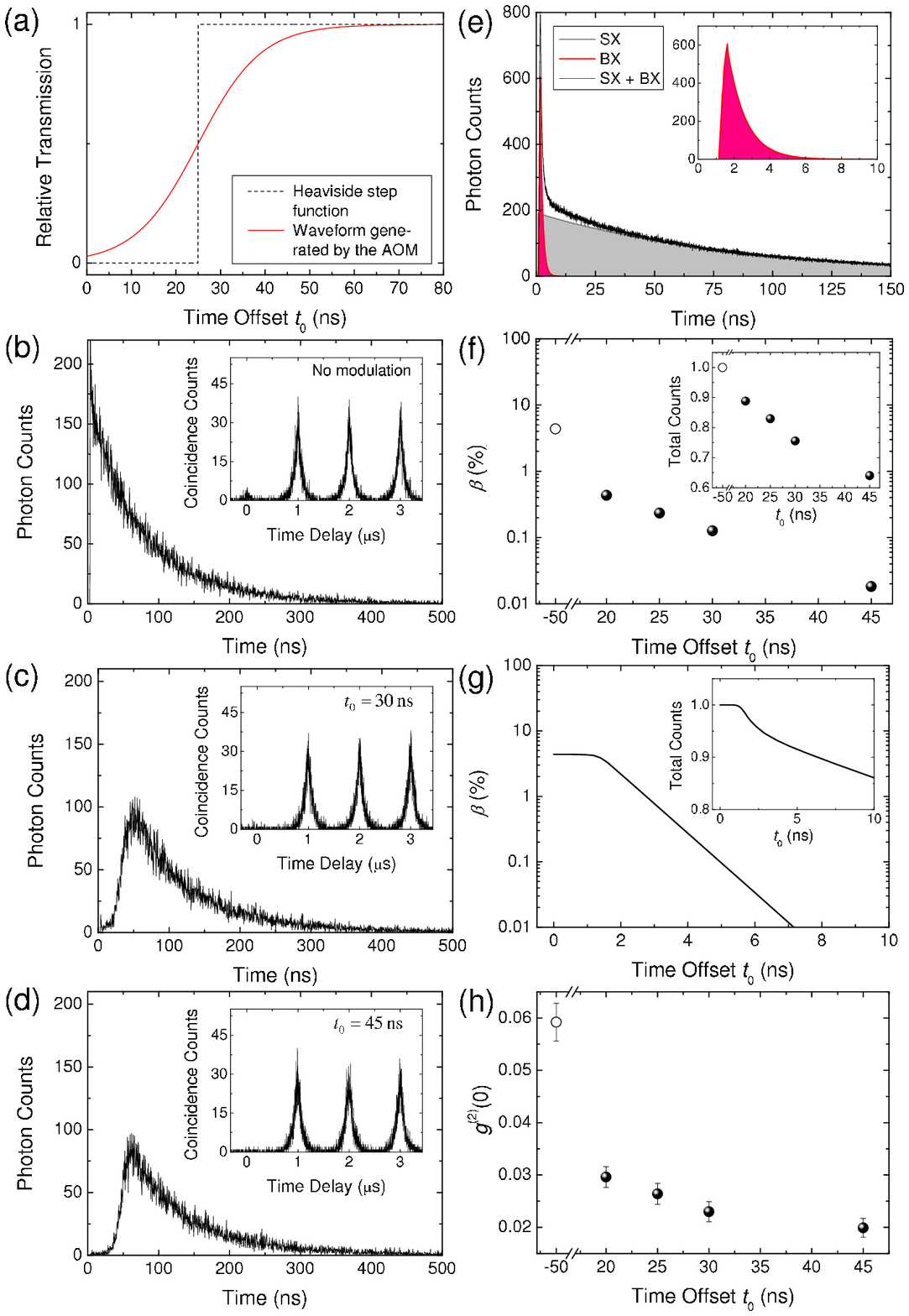}
\caption{\label{fig:3} (color online) Purification of the single photons from room-temperature quantum dots. (a) The modulation function generated by the acousto-optic modulator (red solid curve) compared to the Heaviside step function (black dashed curve). (b) The waveform and Hanbury-Brown-Twiss measurement (inset) of the unmodulated single photons. (c)-(d) The waveforms and Hanbury-Brown-Twiss measurements (insets) of the modulated single photons with time offset of $t_0 = 30$ ns (c) and $t_0 = 45$ ns (d). (e) Decomposition of the photoluminescence (black curve) into single-exciton (SX, pink) and biexciton (BX, gray) emission. (f) The biexciton percentage and relative total counts (inset) with different time offsets. (g) The biexciton percentage and relative total counts (inset) with different time offsets if the modulation function is Heaviside step. (h) The normalized photon correlation functions $g^{(2)}(0)$ at zero time delay. The time offset of the unmodulated single photons, whose values are marked by open circles, corresponds to $t_0 = -50$ ns due to the finite rising time of the modulation waveform.}
\end{figure}

The possibility of manipulating the waveform allows us to purify the single photons. As can be seen in Fig.~\ref{fig:2}(a), the two-photon emission due to the biexciton transition is quite intense in the first few ns, which causes the quality of single photons to degrade. This is also evident in Fig.~\ref{fig:1}(c) where a small peak is noticeable at the time delay ($\tau \sim 0 \ \mu$s) that antibunching occurs. Thus, to purify the single photons, one can superimpose a Heaviside-step function $H(t-t_0)$ onto the unmodulated waveform with a proper time offset $t_0$. This would eliminate the multiphoton emission from the quantum dot. In practice, the modulation waveform is smoothed at the rising edge due to the finite optical rise time of the acousto-optic modulator [Fig.~\ref{fig:3}(a)].

Figure~\ref{fig:3} summarizes the purification of the single photons from a single quantum dot. For comparison, the waveform of the unmodulated single photons is shown in Fig.~\ref{fig:3}(b), the inset of which shows a peak of nonzero correlation in the antibunching region (centered at $\tau = 0 \ \mu$s) of the Hanbury-Brown-Twiss measurement. Figure~\ref{fig:3}(c) and (d) show two examples of the modulated single photons with the corresponding Hanbury-Brown-Twiss measurements given in the insets. The area of the peak centered at $\tau = 0 \ \mu$s reduces noticeably and vanishes with larger $t_0$, indicating that the biexciton emission is completely eliminated. These observations are consistent with the fact that the the biexciton emission occurs only within the first few nanoseconds. 

In Fig.~\ref{fig:3}(e), we estimate the percentage $\beta$ of the photons from the biexciton emission in the absence of waveform modulation. The total emission is assumed to consist of single-exciton and biexciton emission. Noting that the biexciton emission is extinct after a few ns (inset), we reconstruct the time trace of the single-exciton emission (gray area) by fitting the photoluminescence at a later time ($>$~100~ns) with a single exponential decay. The time trace of the biexciton emission (pink area) is then revealed by subtracting that of the single-exciton emission from the total emission. By comparing the areas under the time traces, we estimate that the biexciton emission is $4 \%$ of the total emission. Using the same analysis, we also obtain the biexciton percentage in the presence of modulation. The results are summarized in Fig.~\ref{fig:3}(f) where one can see a significant drop of biexciton percentage after the first few nanoseconds and it being nearly eliminated at larger time offsets. While the biexciton percentage is reduced by the waveform modulation, the total counts of the emitted photons (inset) is also reduced. By using modulators with higher speeds such as electro-optic modulators, the biexciton emission can be temporally eliminated with a higher precision and the reduction of the single-exciton emission can be minimized. This is shown in Fig.~\ref{fig:3}(g) where we calculate the biphoton percentage and the corresponding total count if an ideal Heaviside step waveform is used.

\begin{figure}[t]
\centering
\includegraphics[width=0.9 \linewidth]{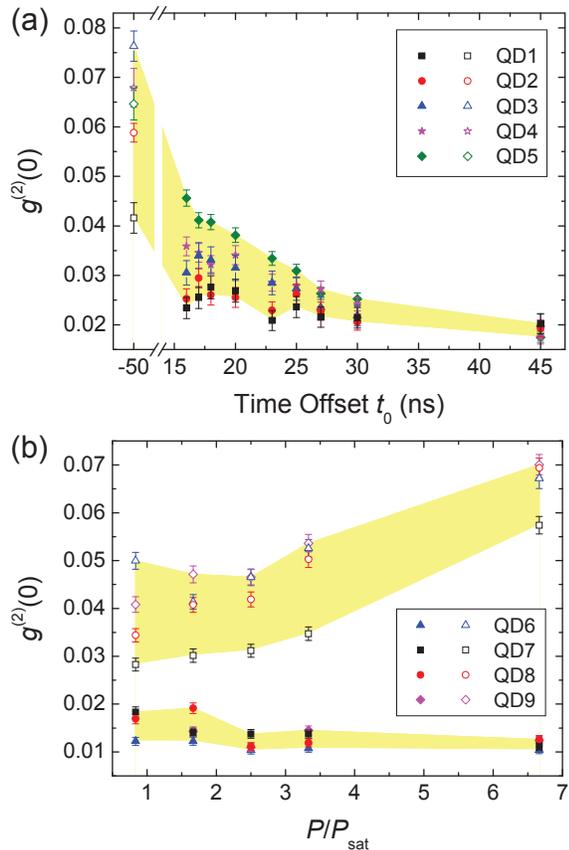}
\caption{\label{fig:4} (color online) Normalized photon correlation function at zero time delay. (a) The time offset of the modulation function is varied from 16 to 45~ns for different quantum dots. (b) The excitation power is varied from 1.4 to 7 $P_{\rm sat}$ for different quantum dots. $g^{(2)}(0)$ of the unmodulated and modulated single photons are shown by the open and solid symbols, respectively.}
\end{figure}

The purity of the waveform-controlled single photons can be characterized by the normalized photon correlation function at zero time delay $g^{(2)}(0)$, which gives an upper bound on the probability of the multiphoton emission~\cite{Santori01}. For pulsed single photon sources, $g^{(2)}(0)$ is equal to the ratio of the peak area at zero time delay ($\tau = 0 \ \mu$s) to the average of the peak areas at nonzero time delays ($\tau \neq 0 \ \mu$s). Figure~\ref{fig:3}(h) shows the $g^{(2)}(0)$ of the single photons before ($t_0 = -50$ ns) and after ($t_0 > 15$~ns) the modulation, with the latter being clearly smaller. The $g^{(2)}(0)$ decreases with $t_0$ in a similar way as the biexciton percentage. The two are approximately related by $g^{(2)}(0) \approx 2 \beta$ \cite{Santori04} if we consider the exciton and biexciton emission only and assume low biexciton percentage. In practice, the dark counts of the detectors, the leakage light of the excitation laser, and the higher-order multiphoton emission can also contribute to $g^{(2)}(0)$.

Figure~\ref{fig:4}(a) shows the $g^{(2)}(0)$ from five different quantum dots at the excitation power of $6.7 P_{\rm sat}$. The $g^{(2)}(0)$ of the unmodulated single photons varies from 0.04 to 0.08. However, with modulation, both the magnitude and variation of $g^{(2)}(0)$ decrease. The different degrees and rates of purifying single photons from different quantum dots indicates the initial variation of the strength and lifetime of the biexciton emission. Nevertheless, by controlling the waveform of the single photons, we achieve $g^{(2)}(0) = 0.02$ for all quantum dots at large time offsets.

%When the excitation power increases, the purity of the unmodulated single photons from a room-temperature quantum dot typically degrades. This is because the probability of the biexciton emission increases with the excitation power. In addition, the single-exciton and biexciton emission at the room temperature are spectrally broad and overlapped. However, with the elimination of the biexciton emission, the waveform-controlled single photons preserve their high purity with increasing excitation power. This is demonstrated in Fig.~\ref{fig:4}(b) which shows the $g^{(2)}(0)$ of the unmodulated (solid symbols) and modulated (open symbols) single photons from three quantum dots for excitation power from 20 to 100~nW. While the $g^{(2)}(0)$ of the unmodulated single photons increases from 0.04 to 0.07, the $g^{(2)}(0)$ of the modulated single photons remains at 0.01 even under higher excitation power.

Figure~\ref{fig:4}(b) shows the $g^{(2)}(0)$ of the single photons from four quantum dots at different excitation power. When the excitation power increases, the probability of the biexciton emission increases and the purity of the unmodulated single photons degrades. The measured $g^{(2)}(0)$ of the unmodulated single photons (open symbols) increases up to $0.06 \sim 0.07$ at an excitation power of $6.7 P_{\rm sat}$. On the other hand, with the biexciton emission eliminated by the modulation, the waveform-controlled single photons preserve the high purity with increasing excitation power. The measured $g^{(2)}(0)$ of the modulated single photons (solid symbols) remains to be $\sim 0.01$ at higher excitation powers.

In summary, we have demonstrated waveform-controlled single photons from single room-temperature quantum dots. By temporally shaping the single photons to eliminate the biexciton emission, single-photon purity of $g^{(2)}(0)=0.01$ is obtained with different quantum dot samples. Among the solid-state SPEs, $g^{(2)}(0)=0.01$ or lower has only been achieved with the cryogenic-temperature InGaAs quantum dots~\cite{Aharonovich16}. Thus, our work provides a novel way of preparing high-purity single photons from room-temperature quantum dots. The purity of the waveform-controlled single photons also remains high and constant with increasing excitation power, thereby allowing one to increase the generation rate without compromising the quality of the single photons. The generation rate will eventually be constrained by the inverse of the single photons' temporal length. As a result, one may need to reduce the temporal length of the single photons (for example, by exploiting the Purcell effect or choosing a different type and size of quantum dots) to optimize the generation rate. In our experiment, fluorescence blinking also causes additional drop in the emission rate. However, with the core-shell engineering~\cite{Nasilowski15}, the blinking could be completely suppressed to further increase the emission rate. Colloidal quantum dots have recently been exploited to demonstrate a cavity-free, broadband approach for engineering the interaction between single photons and quantum emitters \cite{Akimov07}, which finds potential applications such as single-photon transistors~\cite{Chang07} and the long-range coupling of qubits \cite{Chen11}. The temporally shaped single photons may open up new possibilities of controlling such photon-emitter interaction. 

%In addition to the modulator's speed, the fluorescence blinking of colloidal quantum dots also limits the highest achievable exciton emission rate. Nevertheless, with core-shell engineering \cite{Nasilowski15}, the fluorescence blinking can be completely suppressed and 100\% quantum yield of exciton emission can be achieved. 

The authors thank W.-C. Chou, J.-J. Su, W.-H. Chang, Y.-C. Chen, C.-Y. Wang, and C.-H. Wu for helpful discussion and loan equipment. This work was supported by the Ministry of Science and Technology, Taiwan (Grant No. 103-2112-M-007-015-MY3 and No. 106-2112-M-007-023).

S.-W. Feng and C.-Y. Cheng contributed equally to this work.

%\bibliography{flux}

\begin{thebibliography}{99}

\bibitem{Kimble08} H. J. Kimble, Nat. Photonics \textbf{453}, 1023 (2008).

\bibitem{O'Brien09} J. L. O'Brien, A. Furusawa, and J. Vuckovic, Nat. Photonics \textbf{3}, 687 (2009).

\bibitem{Northup14} T. E. Northup and R. Blatt, Nat. Photonics \textbf{8}, 356 (2014).

\bibitem{Shields07} A. J. Shields, Nat. Photonics \textbf{1}, 215 (2007).

\bibitem{Michler00b} P. Michler, A. Imamo$\breve{\rm g}$lu, M. D. Mason, P. J. Carson, G. F. Strouse, and S. K. Buratto, Nature (London) \textbf{406}, 968 (2000).

\bibitem{Lounis00} B. Lounis, H. A. Bechtel, D. Gerion, P. Alivisatos, and W. E. Moerner, Chem. Phys. Lett. \textbf{329}, 399 (2000).

%\bibitem{Cirac97} Cirac, J. I., Zoller, P., Kimble, H. J. \& Mabuchi, H. Quantum State Transfer and Entanglement Distribution among Distant Nodes in a Quantum Network. \textit{Phys. Rev. Lett.} \textbf{78}, 3221--3224 (1997).
%\bibitem{Cirac97} J. I. Cirac, P. Zoller, H. J. Kimble, and H. Mabuchi, Phys. Rev. Lett. \textbf{78}, 3221 (1997).

%\bibitem{Gorshkov07} Gorshkov, A. V., Andr$\acute{\rm e}$, A, Fleischhauer, M., S{\o}rensen, A. S. \& Lukin, M. D. Universal Approach to Optimal Photon Storage in Atomic Media. \textit{Phys. Rev. Lett.} \textbf{98}, 123601 (2007).
%\bibitem{Gorshkov07} A. V. Gorshkov, A. Andr$\acute{\rm e}$, M. Fleischhauer, A. S. S{\o}rensen, and M. D. Lukin, Phys. Rev. Lett. \textbf{98}, 123601 (2007).

%\bibitem{Zhang12} Zhang, S., Liu, C., Zhou, S., Chuu, C.-S., Loy, M. M. T. \& Du, S. Coherent Control of Single-Photon Absorption and Reemission in a Two-Level Atomic Ensemble. \textit{Phys. Rev. Lett.} \textbf{109}, 263601 (2012).
%\bibitem{Zhang12} S. Zhang, C. Liu, S. Zhou, C.-S. Chuu, M. M. T. Loy, and S. Du, Phys. Rev. Lett. \textbf{109}, 263601 (2012).

%\bibitem{Inoue02} Inoue, K., Waks, E., \& Yamamoto, Y. Differential phase shift quantum key distribution. \textit{Phys. Rev. Lett.} \textbf{89}, 037902 (2002).
%\bibitem{Inoue02} K. Inoue, E. Waks, and Y. Yamamoto, Phys. Rev. Lett. \textbf{89}, 037902 (2002).

%\bibitem{Liu13} Liu, C., Zhang, S., Zhao, L., Chen, P., Fung, C. -H. F., Chau, H. F., Loy, M. M. T. \& Du, S. Differential-phase-shift quantum key distribution using heralded narrow-band single photons. \textit{Opt. Express} \textbf{21}, 9505--9513 (2013).
%\bibitem{Liu13} C. Liu, S. Zhang, L. Zhao, P. Chen, C.-H. F. Fung, H. F. Chau, M. M. T. Loy, and S. Du, Opt. Express \textbf{21}, 9505 (2013).

%\bibitem{Keller04} Keller, M., Lange, B., Hayasaka, K., Lange, W. \& Walther, H. Continuous generation of single photons with controlled waveform in an ion-trap cavity system. \textit{Nature} \textbf{431}, 1075--1078 (2004).
\bibitem{Keller04} M. Keller, B. Lange, K. Hayasaka, W. Lange, and H. Walther, Nature (London) \textbf{431}, 1075 (2004).

%\bibitem{Balic05} Balic, V., Braje, D. A., Kolchin, P., Yin, G. Y. \& Harris, S. E. Generation of Paired Photons with Controllable Waveforms. \textit{Phys. Rev. Lett.} \textbf{94}, 183601 (2005).
\bibitem{Balic05} V. Balic, D. A. Braje, P. Kolchin, G. Y. Yin, and S. E. Harris, Phys. Rev. Lett. \textbf{94}, 183601 (2005).

%\bibitem{Du08} Du, S., Kolchin, P., Belthangady, C., Yin, G. Y., \& Harris, S. E. Subnatural Linewidth Biphotons with Controllable Temporal Length. \textit{Phys. Rev. Lett.} \textbf{100}, 183603 (2008).
\bibitem{Du08} S. Du, P. Kolchin, C. Belthangady, G. Y. Yin, and S. E. Harris, Phys. Rev. Lett. \textbf{100}, 183603 (2008).

%\bibitem{Kolchin08} Kolchin, P., Belthangady, C., Du, S., Yin, G. Y. \& Harris, S. E. Electro-Optic Modulation of Single Photons. \textit{Phys. Rev. Lett.} \textbf{101}, 103601-103604 (2008).
\bibitem{Kolchin08} P. Kolchin, C. Belthangady, S. Du, G. Y. Yin, and S. E. Harris, Phys. Rev. Lett. \textbf{101}, 103601 (2008).

%\bibitem{Chen10} Chen, J. F., Zhang, S., Yan, H., Loy, M. M. T., Wong, G. K. L. \& Du, S. Shaping biphoton temporal waveforms with modulated classical fields. \textit{Phys. Rev. Lett.} \textbf{104}, 183604 (2010).
\bibitem{Chen10} J. F. Chen, S. Zhang, H. Yan, M. M. T. Loy, G. K. L. Wong, and S. Du, Phys. Rev. Lett. \textbf{104}, 183604 (2010).

%\bibitem{Zhao15} Zhao, L., Guo, X., Sun, Y., Su, Y., Loy, M. M. T. \& Du, S. Shaping the biphoton temporal waveform with spatial light modulation. \textit{Phys. Rev. Lett.} \textbf{115}, 193601 (2015).
\bibitem{Zhao15} L. Zhao, X. Guo, Y. Sun, Y. Su, M. M. T. Loy, and S. Du, Phys. Rev. Lett. \textbf{115}, 193601 (2015).

%\bibitem{Michler00} Michler, P. Kiraz, A., Becher, C., Schoenfeld, W. V., Petroff, P. M., Zhang, L., Hu, E. \& Imamoglu, A. A Quantum Dot Single-Photon Turnstile Device. \textit{Science} \textbf{290}, 2282--2285 (2000).
%\bibitem{Michler00a} P. Michler, A. Kiraz, C. Becher, W. V. Schoenfeld, P. M. Petroff, L. Zhang, E. Hu, and A. Imamo$\breve{\rm g}$lu, Science \textbf{290}, 2282 (2000).

%\bibitem{Ates13} Ates, S., Agha, I., Gulinatti, A., Rech, I., Badolato, A. \& Srinivasan, K., Improving the performance of bright quantum dot
%single photon sources using temporal filtering via amplitude modulation. \textit{Sci. Rep.} \textbf{3}, 1397 (2013).

%\bibitem{Lounis00} Lounis, B., Bechtel, H. A., Gerion, D., Alivisatos, P. \& Moerner, W. E. Photon antibunching in single CdSe/ZnS quantum dot fluorescence. \textit{Chem. Phys. Lett.} \textbf{329}, 399--404 (2000).

%\bibitem{Mahler08} Mahler, B., Spinicelli, P., Buil, S., Quelin, X., Hermier, J.-P. \& Dubertret, B. Towards non-blinking colloidal quantum dots. \textit{Nat. Materials} \textbf{7}, 659--664 (2008).
%\bibitem{Mahler08} B. Mahler, P. Spinicelli, S. Buil, X. Quelin, J.-P. Hermier, and B. Dubertret, Nat. Materials \textbf{7}, 659 (2008).

%\bibitem{Fisher05} Fisher, B., Caruge, J. M., Zehnder, D. \& Bawendi, M. G. Room-Temperature Ordered Photon Emission from Multiexciton States in Single CdSe Core-Shell Nanocrystals. \textit{Phys. Rev. Lett.} \textbf{94}, 087403 (2005).
\bibitem{Fisher05} B. Fisher, J. M. Caruge, D. Zehnder, and M. G. Bawendi, Phys. Rev. Lett. \textbf{94}, 087403 (2005).

%\bibitem{Brokmann04} Brokmann, X., Coolen, L., Dahan, M. \& Hermier, J. P. Measurement of the Radiative and Nonradiative Decay Rates of Single CdSe Nanocrystals through a Controlled Modification of their Spontaneous Emission. \textit{Phys. Rev. Lett.} \textbf{93}, 107403 (2004).
\bibitem{Brokmann04} X. Brokmann, L. Coolen, M. Dahan, and J. P. Hermier, Phys. Rev. Lett. \textbf{93}, 107403 (2004).

%\bibitem{Santori01} Santori, C., Pelton, M., Solomon, G., Dale, Y. \& Yamamoto, Y. Triggered Single Photons from a Quantum Dot. \textit{Phys. Rev. Lett.} \textbf{86}, 1502--1505 (2001).
\bibitem{Santori01} C. Santori, M. Pelton, G. Solomon, Y. Dale, and Y. Yamamoto, Phys. Rev. Lett. \textbf{86}, 1502 (2001).

\bibitem{Santori04} C. Santori, D. Fattal, J. Vuckovic, G. S. Solomon, and Y. Yamamoto, New J. Phys. \textbf{6}, 89 (2004).

\bibitem{Aharonovich16} I. Aharonovich, D. Englund, and M. Toth, Nat. Photonics \textbf{10}, 631 (2016).

\bibitem{Nasilowski15} M. Nasilowski, P. Spinicelli, G. Patriarche, and B. Dubertret, Nano Lett. \textbf{15}, 3953 (2015).

\bibitem{Akimov07} A. V. Akimov, A. Mukherjee, C. L. Yu, D. E. Chang, A. S. Zibrov, P. R. Hemmer, H. Park, and M. D. Lukin, Nature (London) \textbf{450}, 402 (2007).

\bibitem{Chang07} D. E. Chang, A. S. S{\rm $\o$}rensen, E. A. Demler, and M. D. Lukin, Nat. Phys. \textbf{3}, 807 (2007).

\bibitem{Chen11} G.-Y. Chen, N. Lambert, C.-H. Chou, Y.-N. Chen, and F. Nori, Phys. Rev. B \textbf{84}, 045310 (2011).

\end{thebibliography}

\end{document}